\documentclass [12pt,preprint]{aastex}

\usepackage{epsfig}

\begin{document}
\voffset-1cm
\newcommand{\gsim}{\hbox{\rlap{$^>$}$_\sim$}}

\title{The Supernova associated with GRB 030329}

\author{Shlomo Dado\altaffilmark{1}, Arnon Dar\altaffilmark{1} and
A. De R\'ujula\altaffilmark{2}}

\altaffiltext{1}{dado@phep3.technion.ac.il, arnon@physics.technion.ac.il, 
dar@cern.ch.\\
Physics Department and Space Research Institute, Technion, Haifa 32000, 
Israel}
\altaffiltext{2}{alvaro.derujula@cern.ch.
Theory Division, CERN, CH-1211 Geneva 23, Switzerland}


\begin{abstract}

The relative proximity of the recent gamma ray burst (GRB) 030329
resulted in a large gamma-ray fluence and in the brightest-ever
afterglow (AG), hours after the burst, in the radio, optical and
X-ray bands, permitting precise AG measurements, sensitive tests
of models and an excellent occasion to investigate the association
of GRBs with supernovae (SNe). The Cannonball (CB) model provides
a good, simple and universal description of all AGs of GRBs of
known redshift, so  that it is straightforward to use it to predict
what the expected SN signatures are. In the case of GRB 030329, 10
days after burst the AG should begin to reveal the lightcurve,
spectrum and polarization of an underlying SN ---akin to SN1998bw---
which will peak in the NIR/optical band around day 15.  These
effects will be easily observable if indeed SN1998bw is a good
``standard candle'' for GRB-associated SNe and if the so far unknown
extinction in the host galaxy is not too large.


\end{abstract}


\section*{Introduction}

It is important to know what it is that makes GRBs.
The evidence for an association of long duration GRBs with 
supernovae (SNe)
has been treated by observers with due care, since in only a few
cases is the evidence very clear and model-independent.
Moreover, the only overwhelmingly clear association ---that of
GRB 980425 with SN1998bw--- does not fit at all in the generally
accepted fireball models: the GRB is far too weak and the SN
is annoyingly peculiar. These may be the reasons why
authoritative reviews of the subject, even very recent ones, 
do not consider the SN/GRB association to be an
established fact (e.g. Waxman 2003). To judge from  
recent Gamma Ray Burst Communication Network (GCN)
postings concerning GRB 030329, the observers' attitude is more 
positive: many GCNs discuss minor changes in the smooth
afterglow (AG) decline as possible SN signatures, even much
before ---and a fluence level much larger--- than one may expect them
to occur.

In the CB model long-duration 
GRBs are assumed to be associated with common core-collapse 
SNe (Dar and De R\'ujula 2000). In this model there is nothing
special about GRB 980425, which is describable
in exactly the same way as all other GRBs of known redshift;
nor is SN1998bw particularly peculiar: it did not emit the observed
puzzling radio and X-ray signals, the CB responsible for the
GRB did it, as discussed in tiresome detail in Dado el al. (2003a).
It follows that, {\it in the CB model,} it makes sense to use SN1998bw
as a putative standard candle for SNe associated with GRBs.
The result of that exercise is that, {\it in the CB model,} there is
evidence for a GRB/SN association in all cases in which the SN could
in practice be observed, about 1/2 of the total (Dado et al.~2002a,b,c;
2003a,b,c).

The Cannonball (CB) model provides a good, simple and universal
description of all AGs of GRBs of known redshift, $z$, so that it is
straightforward to use it to predict what the expected SN contribution
to the AG should be, since one is confident on the extrapolated
CB contribution, a ``background'' to the SN signal.
 By making a fit to the observed AGs, {\it
before} the SN contribution became observable, we have, in three
cases, succesfully {\it predicted} what this contribution would be
(Dado et al.~2002b,c, 2003b).  We endeavour to do the same for GRB 
030329, by submitting this paper to the astro/ph archives $\sim\! 2$
days before the SN 
contribution ought to become discernible. The credibility of our
predictions is weakened by the fact that the data, so far preliminary
and unpublished, are insufficient to estimate the 
extinction in the host galaxy. Also if ---unlike in all other cases for 
which there is evidence of an association--- the SN that generated 
GRB 030329 turns out to be more than two times less 
intrinsically luminous than SN1998bw, it will be difficult to
see its contribution to the AG light curves, though its presence
may still significantly distort the late-time AG spectrum.

We also show that --in the CB model-- most of the
``unusual'' features of the AG of GRB 030329 occured before in other GRBs 
and could, in this sense, be anticipated. 
The early optical AG of GRB 030329 is, as for GRBs 990123, 021004 and
021121, a direct tracer of the expected circumburst density profile:
in all these cases 
the AGs were ``caught'' very early and they
provide extremely clean evidence for the expected
$1/r^2$ profile generated by a pre-SN stellar wind (Dado et al. 2003c).
Both GRB 030329 and 021004 have $\gamma$-ray light curves 
dominated by two pulses or, in the CB model, two cannonballs.
Both AGs display two wide shoulders that, in the same model,
correspond to the contributions of the two CBs.
The superimposed variations in the predicted smooth
AG lightcurve, ups and downs of $\sim 1/2$ magnitude,
 are, as for GRBs 021004, 000301c and 970508, to be
expected:  they directly trace, like a needle on a vynil record,
moderate deviations from a constant-density interstellar medium.

\section{GRB 030329}

The bright optical AG of this GRB
(Vanderspek et al.~2003), which
 was first observed $\sim\!1.25$h after burst
by Peterson and Price (2003) and by Torii (2003) is ---at magnitude $R\! \sim\!
12.5$--- by far the brightest optical afterglow observed hours after
burst.   The X-ray AG, also very bright, was seen at time $t\!\sim$ 5h 
after burst with RXTE,
by Marshall and Swank (2003). The bright radio AG was observed at
$t\!\sim\!14$h by Berger et al.~(2003). The redshift of the host
galaxy was first determined by Greiner et al. (2003) to be $z=0.1685$. 
The large luminosity of this AG has triggered a large
interest, reflected in a copious and enthousiastic release of GCN
communications.

\section*{The Cannonball Model of GRBs}

In the CB model (Dar and De R\'ujula 2000, 2001; Dado et al. 2002a; 2003a)
reviewed in De R\'ujula 2002 and Dar 1993), long duration GRBs and
their AGs are produced in core-collapse supernovae akin to SN1998bw by the
ejection of bipolar jets of ordinary-matter, hydrogenic plasma clouds or
``cannonballs'' (CBs) with high Lorentz factors ($\gamma_0\sim 10^3$).  A CB
is emitted, as observed in $\mu$-quasars, when part of an accretion disk
falls abruptly onto the newly-born compact central object. Crossing the
circumburst shells with a large $\gamma_0$, the surface of a CB is
collisionally heated to keV temperatures and the thermal radiation it
emits as it reaches the transparent outskirts of the shells ---boosted and
collimated by the CB's motion--- is a single $\gamma$-ray pulse in a GRB.
The cadence of pulses reflects the chaotic accretion and is not
predictable, but the individual-pulse temporal and spectral properties are
(Dar and De R\'ujula 2001;  Dar 2003). In practice GRBs
are observable only if the angle $\theta$ subtended by the CBs' velocity
vector and the line of sight to the observer is small:
$\theta={\cal{O}}(1/\gamma_0)$. 

After becoming visible, a CB first cools by bremsstrahlung and expansion,
emitting a hard spectrum that is seen in the X-ray band with a fluence
decreasing with the expected $1/t^5$ behaviour.  
When a CB's temperature approaches $ \sim 1$ eV (within a few observer
minutes), its emissivity is dominated by synchrotron emission from the
electrons that penetrate in it as it propagates in the interstellar medium
(ISM).  Integrated over frequency, this synchrotron emissivity is
proportional to the energy-deposition rate of the ISM electrons in the CB.
These electrons are Fermi accelerated in the CB's tangled magnetic maze to
a broken power-law energy distribution with a ``bend'' energy equal to
their incident energy in the CBs' rest frame. Their synchrotron radiation
---the afterglow--- is also  collimated and
Doppler-boosted by the relativistic motion of the CBs. 
The radiation is also redshifted by the
cosmological expansion and attenuated 
on its
way to an earthly observer, during its passage through the CB
itself, the host galaxy, the intergalactic space and our own galaxy.

\section{The NIR/Optical AG in the CB model}

In the CB model, the observed AGs have three origins: the ejected CBs, the
concomitant SN explosion, and the host galaxy. These components are
usually unresolved in the measured ``GRB afterglows'', so that the
corresponding light curves and spectra are the cumulative energy flux
density:
\begin{equation}
    F_{AG}=F_{CBs}+F_{SN}+F_{HG}\, .
\label{sum}
\end{equation}
The contribution of the host galaxy, $F_{HG}$, is usually extracted from
``very'' late time observations when the CB and SN contributions become
negligible, or is best fitted if such data are not available. 

Let the energy flux
density of SN1998bw at redshift $ z_{bw}=0.0085$ (Galama et al. 1998) 
be $ F_{bw}[\nu,t]$. For a similar SN placed at a redshift $ z$:   
\begin{eqnarray}
{ F_{SN}[\nu,t] = } && {{1+z \over 1+z_{bw}}\;
{D_L^2(z_{bw})\over D_L^2(z)}}\, \times\nonumber \\ &&
{
F_{bw}\left[\nu\,{1+z \over 1+z_{bw}},\;t\, {1+z_{bw} \over 1+z}\right]\;   
A_{SN}(\nu,z)}\, ,
\label{bw}
\end{eqnarray}
where $ A_{SN}(\nu,z)$ is the attenuation along the line
of sight and $ D_L(z)$ is the
luminosity distance (we use a cosmology with
${ \Omega_M}=0.3$, ${
\Omega_\Lambda}=0.7$ and  $ H_0=65$ km/s/Mpc).

In the rest frame of a CB the AG that an observer sees at optical
frequencies is given by:
\begin{equation}
 F_{_{CB}}[\nu,t]=
{f\, [\gamma(t)]^2\, n_p\, \over \nu_b}{[\nu/\nu_b]^{-1/2}\over
\sqrt{1+[\nu/\nu_b]^{(p-1)}}}\; ,
\label{fluxdensity2}
\end{equation}
where $ f$ is a normalization constant (see Dado et
al.~2003a for its theoretical estimate), $ \gamma(t)$ is the Lorentz
factor of the CB, $ p\approx 2.2 $ is the spectral index of the radiating 
electrons in the CB, and $ \nu_b$ is the ``injection bend'' frequency
for an interstellar density $ n_p$:
\begin{equation}
 \nu_b \simeq 1.87\times 10^3\, [\gamma(t)]^3\,
\left[{n_p\over 10^{-3}\;cm^3}\right]^{1/2}\, Hz.
\label{nubend}
\end{equation}
The theoretical motivation, as well as the excellent observational support
for this ``bend'', are discussed in Dado et al. 2003a.
An observer in the GRB progenitor's rest system,
viewing a CB at an angle $\theta$, sees its radiation
Doppler-boosted by a factor $\delta$:
\begin{equation}
 \delta(t)\equiv
{1\over\gamma(t)\,(1-\beta(t)\cos\theta)}
\simeq {2\,\gamma(t)\over 1+\theta^2\gamma(t)^2}\; ,
\label{doppler}
\end{equation}
where the approximation is valid in the domain of interest for GRBs:
large $\gamma$ and small $\theta$.
The cannonballs' AG spectral energy density $ F^{obs}_{CB}$
seen by a cosmological observer at a redshift $ z$ is:
\begin{equation}
 F^{obs}_{CB}[\nu,t]\simeq
 {A(\nu,t)\, (1+z)\,\delta(t)^3
                    \over 4\, \pi\, D_L^2}\,
F_{_{CB}}\left[{(1+z)\,\nu\over\delta(t)},{\delta(t)\,t\over 1+z}
\right]\! ,
\label{Fnuobser}
\end{equation}
where $A(\nu,t)$ is the correction for the total extinction
of the CB's radiation.

For an interstellar medium of constant baryon density $ n_p$, the   
Lorentz factor  $\gamma(t)$ is given by:
\begin{eqnarray}
 \gamma&=&\gamma(\gamma_0,\theta,x_\infty;t)
= {B^{-1}} \,\left[\theta^2+C\,\theta^4+{1/C}\right]\nonumber\\
 C&\equiv&
\left[{2/
\left(B^2+2\,\theta^6+B\,\sqrt{B^2+4\,\theta^6}\right)}\right]^{1/3}
\nonumber\\
 B&\equiv&
{1/ \gamma_0^3}+{3\,\theta^2/\gamma_0}+
{6\,c\, t/ [(1+z)\, x_\infty]}
\label{cubic}
\end{eqnarray}
where $\gamma_0=\gamma(0)$, and
$ x_\infty\equiv N_{CB}/(\pi\, R_{max}^2\, n_p)$
characterizes the CB's slow-down in terms of
$ N_{CB}$: its baryon number, and $ R_{max}$:
its asymptotic radius (it takes a distance $ x_\infty/\gamma_0$ for
the CB to half its original Lorentz factor).

For AGs that are observed very early (those of the GRBs
021211, 990123, 021004 and 030329) the approximation of a constant
ISM density, adopted above, is not good,
since the CBs are first observed while travelling in the progenitor
star's wind. The total wind grammage is insufficient to
make $\gamma(t)$ deviate from $\gamma_0$ during the
CB's traversal of the wind. Thus the only modification is
that the density occuring in Eqs.~(\ref{fluxdensity2}) and
(\ref{nubend}) now becomes, for a typical ``windy'' profile:
\begin{equation}
n(r)=n_p\, (1+r_0^2/r^2),
\label{windy}
\end{equation}
where $n_p$ is again a constant. These fits have two
extra parameters ($ n_p$, which was
previously embedded with $ f$, or played a very marginal
role via Eq.~(\ref{nubend}), and $r_0$).
Note that at very early time,
 $r\approx c\, t\, \gamma_0\, \delta_0/(1+z)
\propto t\, $ and the above circumburst  wind profile yields
 $F_\nu\propto [1+(\bar{t}/t)^2]^{0.75}$  
(Dado et al. 2003c), 
where $\bar{t}$ is the observer time when $r=r_0$). This very simple
expression fits like the proverbial glove the data 
on the early time behaviour of the AGs of the GRBs
021211, 990123, 021004 and 030329, that
in the fireball models would be attributed to a ``reverse'' shock.

The contribution to $A(\nu,z)$ in Eq.~(\ref{Fnuobser})
from selective extinction in the host
galaxy and in the intergalactic medium can be estimated from the difference
between the observed spectral index {\it at very early time when the CBs
are still near the SN} and that expected in the absence of extinction.
Indeed, the CB model predicts ---and the data confirm with precision---
the gradual evolution of the effective optical spectral index towards the
constant value $\approx -1.1$ observed in all ``late'' AGs 
(Dado et al. 2002a, 2003a). 
The ``late'' index is independent of the attenuation in the host
galaxy, since at $ t>1$ (observer's) days after the explosion, the CBs
are typically already moving in the low column density, optically
transparent halo of the host galaxy.

The comparison in Dado et al.~(2002a,b,c; 2003a,b,c) of the predictions of
Eq.~(\ref{Fnuobser}) with the observations of optical, X-ray and radio
light-curves and spectra for {\it all} GRBs of known redshift are very
simple, satisfactory and parameter-thrifty.

\section{GRB 030329 in the CB model}
\noindent
Six properties of AGs in the CB model are particularly 
relevant to GRB 030329:
\begin{itemize}
\item{}
The optical AG is ``caught'' at very early times, when the CBs are
still crossing the parent star's wind-generated density profile.
\item{}
In high-precission data, density inhomogeneities 
along the CBs' trajectory ---expected within star formation regions and
upon exit from the superbubbles where most SNe take place---
result in observable achromatic ``bumps'' in the AG. 
These features have been seen in GRBs 970508, 000301c 
and 021004  (Dado et al. 2002a, 2003b).
\item{} Except for GRB 021004,  the past AG data were 
course enough or started late enough for
the contributions of different CBs (which can often be resolved as
individual pulses in the GRB phase) to coalesce into an AG 
describable by a single CB or a collection of similar ones.  But
individual CBs may have different properties, or be emitted
at somewhat different angles,
as observed in the $\mu$-quasar  SS 433 (Margon 1984).
\item{}
The predicted AG spectra and their evolution, in particular  the
steepening at the time-varying frequency $\nu_b(t)$ of Eq.(\ref{nubend}) 
towards $\nu^{-1.1}$ at late time,
is very well supported by the data (Dado et al.~2002a, 2003a).
\item{}
The excess polarization of AGs above that induced by the
ISM in the Galaxy may be largely due to
the host galaxy's ISM. In that case, it should be correlated with the
extinction in the host and decline with time as the CBs  move into its 
halo.
\item{}
The $\gamma$-ray light curve of GRB 030329, like that of GRB 021004, shows 
two prominent pulses  (Vanderspek et al. 2003;
http:// space.mit.edu/ HETE/ Bursts /GRB030329/).
\end{itemize} 

In the CB model the two $\gamma$-ray pulses of the last item  
correspond to two dominant CBs.  The good quality 
and early start of the optical data for
GRB 030329 implies that the individual contributions of
the two CBs are discernable in the AG: we must 
fit the broad-band AG light curves with the additive contributions
of {\bf two CBs}, emitted almost in the same direction,
but with otherwise free
parameters (normalization, $\gamma_0$ and $ x_\infty$). We fix the
spectral index $p$ in Eq.~(\ref{fluxdensity2}) to the theoretically expected 
$p=2.2$, assume the density profile given by Eq.~(\ref{windy}),
 and fit simultaneously all the reported
NIR, optical and radio data.

The individual CBs' contribution
are given by Eq.~(\ref{Fnuobser}) which implicitely uses all equations
from to (\ref{fluxdensity2}) to (\ref{cubic}). The free parameters
are the radius $r_0$ at which the wind and ISM densities
are equal and, for each CB,
the viewing angle,  $\theta$, the normalization constant
$ f\times n_p$,
the initial Lorentz factor, $\gamma(0)$ and the decelaration
parameter, $ x_{\infty}$. 

The $R$-band magnitude of a SN akin to 1998bw, displaced to $z=0.1685$,
is $R\!\sim\! 20.2$ at peak brightness. Thus, in
our fits, we neglect the smaller contribution of the host galaxy,
$R\!>\!23.1$ at $2\sigma$ confidence level (Blake and Bloom,
2003; see also Wood-Vasey et al. 2003).

\subsection{NIR-Optical AG} 

\begin{figure}[]
\vskip -1.5cm
\hskip 2truecm
\vspace*{- .4cm}
\plotone{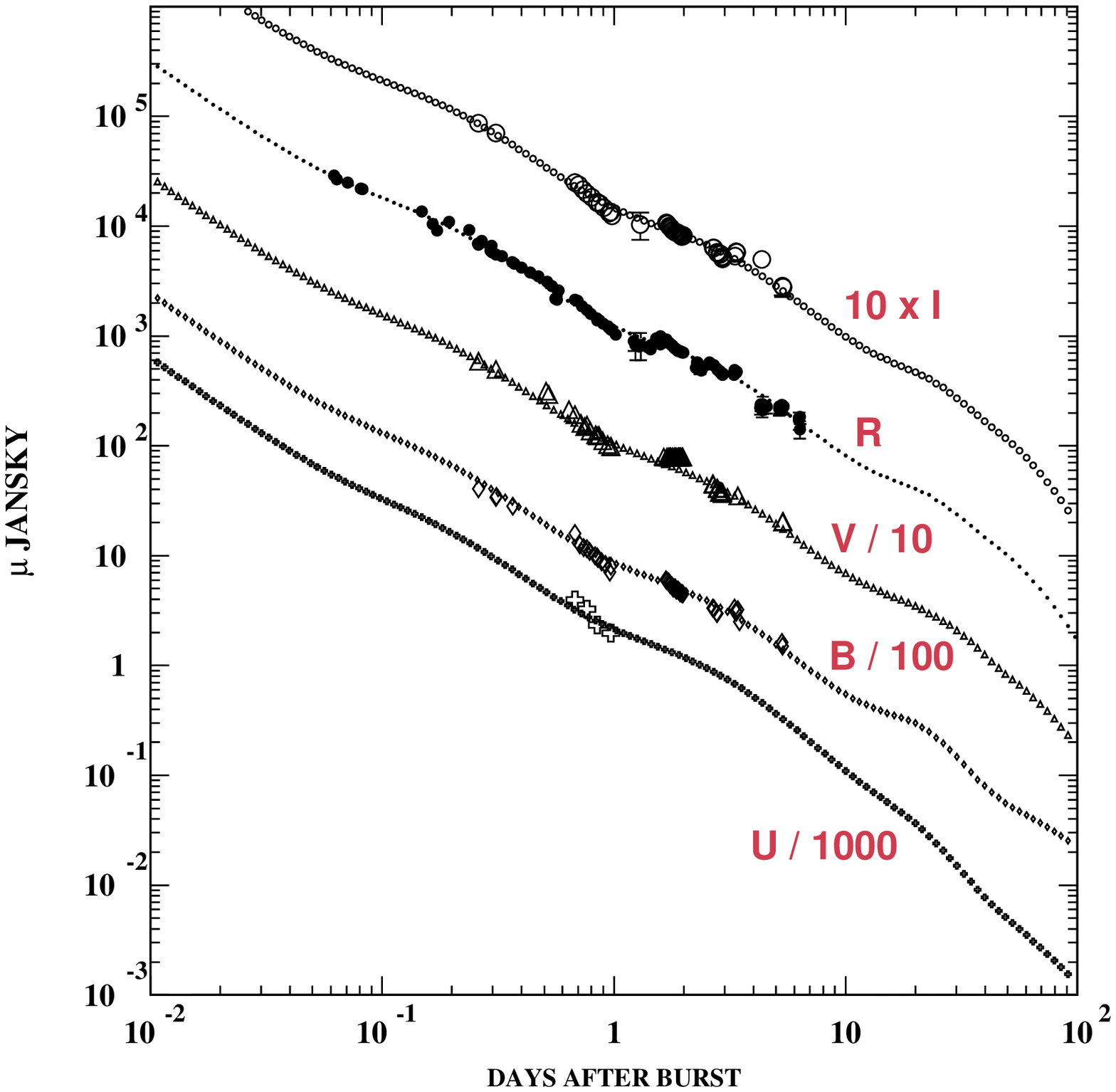}
\figcaption{The NIR--optical observations of the AG of GRB 030329
and the fit for two CBs with different parameters, as described 
in the text. The ISM density is a 
constant plus a ``wind'' contribution decreasing as $ 1/r^2$.  The 
various bands
are scaled for presentation.  The data was selected from the
GCN notices quoted in the text,
recalibrated with the observations of Henden et
al. 2003. The  host-galaxy's contribution
was neglected.  The individual bands have been rescaled for
clarity.
\label{figone}}
\end{figure}

In Fig.~\ref{figone} we show
the CB-model's fit to  data for the NIR-optical light-curves 
reported in the GCN 
circulars 
1990, 2001, 2024, 2046, 2051, 2054, 2079 (Burenin et al.); 
1991, 2005, 2028 (Rumyantsev et al.);
1992 (Andersen et al.);
1995 (Rykoff et al.); 
1999 (Gal-Yam et al.); 
2002, 2035, 2091 (Lipunov et al.); 
2012 (Martini et al.);
2016 (Masi et al.);
2019 (Smith et al.); 
2021 (Halpern et al.); 
2022, 2075 (Zharikov et al.);
2029 (Klose et al.); 
2030 (Bartolini et al.);  
2034, 2045, 2049, 2060 (Lipkin et al.); 
2036 (Garnavich et al.);
2037 (Moran et al.);
2040, 2096 (Lamb et al.); 
2041 (Stanek et al.); 
2042 (Schaefer. et al.); 
2048 (Zeh et al.) 
2056, 2065, 2070 (Fitzgerald \& Orosz); 
2063 (Weidong Li et al.);
2066 (Tober et al.);
2050, 2067, 2083, 2097 (Pavlenko et al.);   
2068 (Bloom et al);    
2074 (Cantiello et al.);  
2077, 2084 (Ibrahimov et al.);
2080 (Sato et al.);
2094 (Khamitov et al.);
2096 (Lee et al.);
and 2098 (Ibrahimov et al.).

All the observational data was recalibrated to the field photometry of Henden
et al.~(2003), either by the observers or by ourselves.   We
have corrected for selective extinction in our galaxy  (Schlegel et al.~1998): $E(B-V)=0.025$ in the direction of GRB
030329. We did not correct for extinction in the host galaxy; the preliminary 
data in the GCN does not provide clear evidence for it.

The broad-band fitted parameters of the two CBs are $\theta[1]=2.00$ mrad,
a nearly identical $\theta[2]=1.95$ mrad for the second CB's direction,
$\gamma_0[1]=1477$,  $\gamma_0[2]=976$ (implying
$\delta_0[1]=306$, $\delta_0[2]=423$), $\rm x_\infty[1]=476$ kpc, and $\rm
x_\infty[2]=36$ kpc.
The density and ``wind'' parameters of Eq.~(\ref{windy}) are
 $n_p=0.86\times 10^{-2}$ cm$^{-3}$ and $r_0=26$ pc, which yield 
$\rho\, r^2=9.3\times 10^{13}$ g  cm$^{-1}$ for the
progenitor-wind's grammage, similar to the canonical value for 
the winds of the parent 
massive stars of core-collapse SNe (Dado et al. 2003c
and references therein).

\begin{figure}[]
\hskip 2truecm
\vspace*{0.8cm}
\vspace*{-.4cm}
\plotone{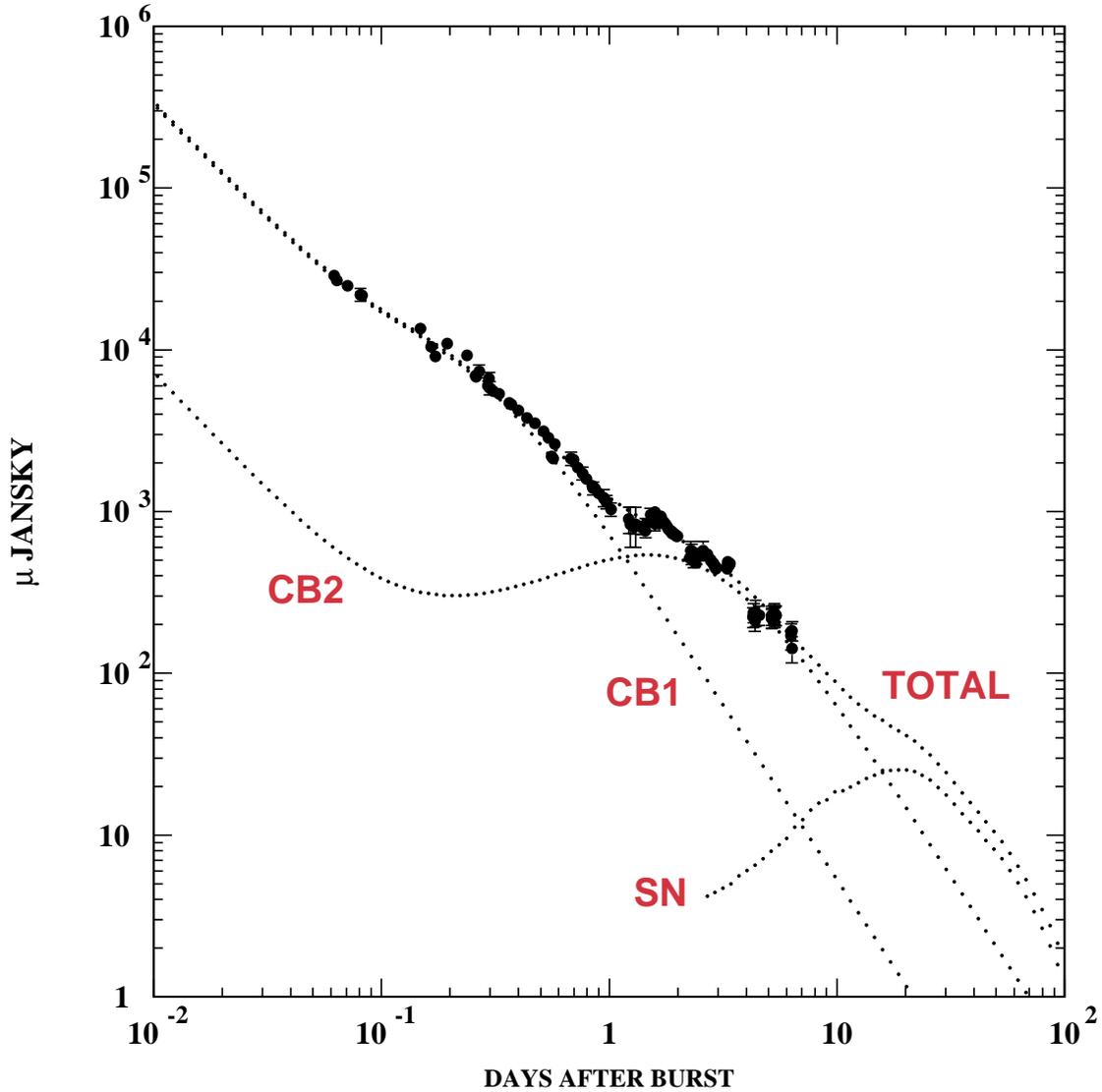}
\figcaption{Blow-up of the R-band results
of Fig. \ref{figone}.
The ISM density was
assumed to be a constant plus an additional ``wind'' contribution
decreasing as $1/r^2$. The wind contribution is
only significant at $\rm t\!<\!0.1$ days, after which the
CBs are more than 10 pc away from the progenitor.
This ``wind'' is also seen in other AGs observed early enough
(Dado et al. 2003c). The individual contributions of the two CBs
and of a SN akin to SN1998bw (at the GRB's redshift) are also shown.
\label{figtwo}}
\end{figure}

To demonstrate the real quality of the fit, we have blown up the R-band
results in Fig.~\ref{figtwo}.  In the region between $t\!\sim\! 1$ and
$\sim\! 5$ days, the data ``wiggles'' by as much as 30\% around the
smoother
theoretical curve.  It would be easy to correct for this by assuming
similar deviations of the ISM density input in Eqs.~(\ref{fluxdensity2})
and (\ref{windy}), relative to a constant $n_p$, clearly a moot
exercise. 

A SN1998bw-like contribution, as
can be seen in Figs.~\ref{figone} and \ref{figtwo}, will be observable.
Since the contribution of the
second CB is still considerable near the peak brightness of the SN 
(at $t\!\sim 15$ days) the SN contribution appears as a shoulder in
the light curves. 
If this
contribution is much fainter ---due to extinction in the host galaxy or a
possible deviation from a so-far-succesful SN1998bw ``standard candle''
ansatz--- its presence could still be
established by the change of colours from the broad band $F_\nu\sim
\nu^{-1.1}$ behaviour of the CBs' emission
towards $F_\nu\sim \nu^{-3.5\pm 1.0}$, the SN spectrum. 

\subsection{Optical Polarization} 

In the CB model, after subtraction of the Galaxy's contribution,
the polarization of
the optical AG is mainly due to the effect of the host's ISM.
It is therefore correlated with the extinction in the host, has a fixed
angular position, and diminishes with time as the CBs escape into the
galactic halo. Consequently near peak brightness of the associated SN
(around 15 days after burst) the polarization of the AG should be that of
the underlying SN.


\subsection{The GRB proper} 

In the CB model, due to Doppler boosting and
relativistic collimation (Dar and De R\'ujula 2000),
the $\gamma$-ray fluence 
of a GRB viewed at a small $\theta$ is
amplified by a huge factor $\delta_0^3$: 
\begin{equation}
F_{_{GRB}}={(1+z) \,\delta_0^3\over 4\,\pi\, D_L^2}\, E_\gamma\, , 
\label{fluence}
\end{equation} 
relative to $E_\gamma$, the total energy in photons emitted by
the CBs in their rest system. The total ``equivalent spherical'', or would-be
isotropic energy, $E^{iso}$, inferred from the observed fluence, is a
factor $(\delta_0)^3$ larger than $E_\gamma$. In Dado et al.~2002a we
deduced that the $E_\gamma$ values of the GRBs of known $z$ span the
surprisingly narrow\footnote{GRBs in the CB model 
are much better standard
candles than in the standard model (Frail et al. 2001).} interval
$10^{44\pm 0.3}$ erg, the spread in $F_{_{GRB}}$ being mainly due to the
spread in their values of $\delta_0$ (deduced 
from the fits to their AGs).
For GRB 030329 the CB-model fit to its broad band AG yields
$\delta_0[1]^3=2.86\times 10^7$ 
and $\delta_0[2]^3=7.56\times 10^7$.
The CB-model expectation from the fitted AG is
$E^{iso} \!\approx\! (\delta_0[1]^3+\delta_0[1]^3)\, E_\gamma\approx 1.04
\times 10^{52\pm 0.3}$ erg, in agreement with the observed $E^{iso}\approx
1.1\times 10^{52}$ erg, deduced from its measured redshift, $z=0.168$,
and fluence in the 30--400 keV band $\sim 10^{-4}$
erg cm$^{-2}$ (Vanderspek et al. 2003).

At the very early stage of $\gamma$-ray emission, CBs are still dense
enough for the ambient protons to interact hadronically with those
of the CB's ordinary-matter plasma.
In the simplest ``surface model'' of the generation of the GRB proper,
the energy deposited in the CBs by these interactions is reemitted
with a quasi-thermal spectrum with a predictable time-dependent
temperature, leading to the correct predictions of the
very stable characteristic energy of GRBs, and of the observed
spectral indices of their ``Band'' spectrum
(Dar and De R\'ujula 2000, Dar 2003).  

The light curve of a GRB's
single pulse rises abruptly as the column density of the material
in front of a CB descends to ${\cal{O}}(1)$ attenuation lengths, and
diminishes thereafter as the density of the material the CB encounters
diminishes with time.  For a windy $1/r^2$ density profile of this density, 
the shape of a pulse at fixed energy is analytical and simple:
\begin{eqnarray}
F_{CB}(t)&\propto& {1\over t^2}\; Exp \left[-2\,{\Delta t\over t}\right],
\nonumber\\
\Delta t&=& {\Delta x \,(1+z)\over c\,\gamma_0\,\delta_0},
\label{shape}
\end{eqnarray}
with $t$ the observer's time and
$\Delta x$ the proper distance travelled by the CB
from the onset of the pulse to the position at which one
$\gamma$-ray absorption length is still in front of it.
In the rough approximation of an 
energy-independent $\gamma$-ray attenuation in the circumburst
material, the shape of an energy-integrated pulse is the same
as in Eq.~(\ref{shape}). 

\begin{figure}[]
\hskip 2truecm
\vspace*{-18cm}
\vspace*{-.4cm}
\plotone{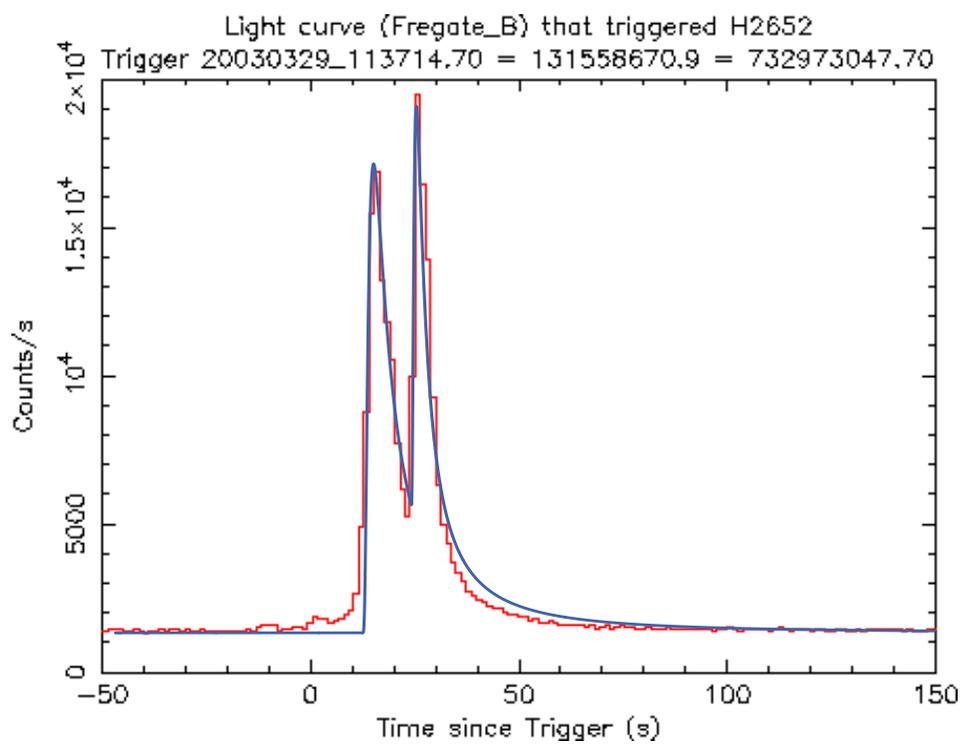}
\figcaption{The $\gamma$-ray light curve of GRB 030329 (the red
binned curve), and its simple CB-model description (the blue continuous
line).
\label{GRB}}
\end{figure}

We have constructed a rough description of the 
$\gamma$-ray light curve of GRB 030329 
(Vanderspek et al. 2003;
http:// space.mit.edu/ HETE/ Bursts /GRB030329/) 
by summing two 
pulses, both with the shape of Eq.~(\ref{shape}), separated by 12
observer's seconds, and with $\Delta t[1]\!=\!6$s, $\Delta t[2]\!=\!4$s.
The result is shown in Fig.~\ref{GRB}. The description of the
GRB is quite satisfactory, even though we would have expected
$\Delta t[1]/\Delta t[2]=
\gamma_0[2]\,\delta_0[2]/(\gamma_0[1]\,\delta_0[1])\!\sim\! 1.1$, as opposed
to 1.5. It would be easy to find excuses to exit from this minor problem
(e.g. we have not quoted errors in our parameters, that are made 
somewhat meaningless
by the slightly erratic behaviour of the AG light curves).

\section{Conclusions}

We have shown that the data on GRB 030329 and on its early NIR-optical
AG are well described by the CB model. This is also the case for all
other GRBs of known redshift, including GRB 980425, whose associated
SN we use as a putative standard candle. 

In the CB model, we contend,
the association between long-duration GRBs and SNe is extremely
well established: in
all GRBs of known redshift (all those with $z\!<\!1.12$)  a
SN1998bw-like supernova could be seen, and 
with various degrees of significance, it was seen. SN1998
stood the ``standard-candle'' test with surprising precision. 
 In many cases, though, the test
depended on very significant extinction corrections in the host
galaxy. If this extinction turns out to be significant for GRB 030329,
the direct observation of a SN signal in the AG could be jeopardized. 
In such a case one would need precise
colour photometry and spectroscopy with the most powerful telescopes, such
as VLT, Magellan, Keck and HST, and very late time measurements of the host
galaxy's magnitude.

{\bf Acknowledgment:} This research was supported in part by the
Helen Asher Space Research Fund at the Technion. Arnon Dar thanks the 
TH division at CERN for its hospitality.

\end{document}